\documentclass{moriond}
\pdfoutput=1

\def\be{\begin{equation}}
\def\ee{\end{equation}}
\def\bea{\begin{eqnarray}}
\def\eea{\end{eqnarray}}

\newcommand{\Photo}{}

\begin{document}
\vspace*{4cm}
\title{CAN WE REACH THE SCALE OF NEW PHYSICS BEHIND THE $B$ ANOMALIES?~\footnote{Contribution~\cite{AGY} to the proceedings of the 2018 Rencontres de Moriond.}}

\author{ TEVONG YOU }

\address{DAMTP, University of Cambridge, Wilberforce Road, Cambridge, CB3 0WA, UK; \\
Cavendish Laboratory, University of Cambridge, J.J. Thomson Avenue, \\ 
Cambridge, CB3 0HE, UK}

{\tiny Cavendish-HEP-18/10, DAMTP-2018-18}

\maketitle\abstracts{
Indirect signs of new physics beyond the Standard Model may be appearing in $B \to K^{(*)}\mu^+\mu^-$ decays. If confirmed, the title question will be of paramount importance in determining the strategy for future colliders. We answer it by estimating the sensitivity to minimal, anomaly-compatible $Z^\prime$ and leptoquark models at the high luminosity LHC, 27 TeV HE-LHC, and 100 TeV FCC-hh; this conservative analysis outlines an upper bound on the available parameter space and the conditions for a reasonable guarantee of direct discovery. 
}

\section{Introduction}

Various measurements involving flavour-changing neutral currents in $B \to K^{(*)}\mu^+\mu^-$ decays are in tension with Standard Model (SM) predictions. Taken individually, none are statistically significant enough yet to claim new physics, but collectively they are remarkably consistent with each other: in a global fit, the pull on a Wilson coefficient is over $4 \sigma$ away from the SM value. The $B$ anomalies are reviewed elsewhere in these proceedings~\cite{Hiller}; it goes without saying that more data and further corroboration by related measurements or independent experiments will be needed to confirm them. Here, let us consider the implications if they are indeed established beyond any reasonable doubt---we will then be in the tantalising position of knowing that fundamentally new physics awaits us at potentially accessible energies. The crucial question is: how accessible exactly? Will a 27 TeV high energy upgrade of the LHC (HE-LHC) be sufficient? Or must we go to 100 TeV as proposed for the FCC-hh? Even then, could the new physics remain just out of reach? 

Ideally, we would have a no-lose theorem for going to higher energy, in the same way that at the LHC the Higgs boson or some other discovery was guaranteed to appear below 1 TeV (or, going back further, how massive electroweak gauge bosons were inevitable on the basis of Fermi's theory). Unfortunately, a similar partial wave analysis of the effective operator involved in $b \to s \mu^+\mu^-$ with the required size of new physics places an upper limit on the scale of unitarity violation at $\sim 80$ TeV~\cite{LM}---beyond the reach of the FCC-hh. Nevertheless, there are still good prospects for discovery given that new physics would typically show up well below the scale of unitarity violation, and other bounds can come into play long before then. 

Our strategy~\cite{AGY} is therefore to consider minimal, pessimistic models of $Z'$ and leptoquarks that UV-complete the effective operator at tree level. We make a first quantitative estimate of the sensitivity to these models at the 27 TeV HE-LHC and 100 TeV FCC-hh. The minimal models are na\"{i}ve in that they include only the couplings necessary to explain the anomalies and are pessimistic in the sense that more realistic models will generally (and sometimes necessarily) induce further couplings to other generations of quarks and leptons that would increase their discoverability potential; demonstrating sensitivity to the pessimistic case therefore implies sensitivity to more realistic, model-dependent cases as a corollary. This forms an alternative way of establishing a more conditional ``no-lose'' theorem (although unlike the unitarity violation limit our bound may still be evaded by even more contrived models).

\section{Future Collider Sensitivity to $Z^\prime$ and Leptoquarks}

We use an approximate method~\cite{TTW} to extrapolate the 95 \% CL limits from current searches for $Z'$ and leptoquarks at the LHC to higher luminosities and energies. It assumes that the limit at a particular mass is mainly driven by the number of background events in a narrow width window, so that the same limit will also apply at higher energy and luminosity at the equivalent mass with the same number of background events. There are clearly limitations to this assumption, but this conservative approach is sufficient for an initial order of magnitude estimate. 

\begin{figure}
\centering
\includegraphics[width=0.39 \textwidth]{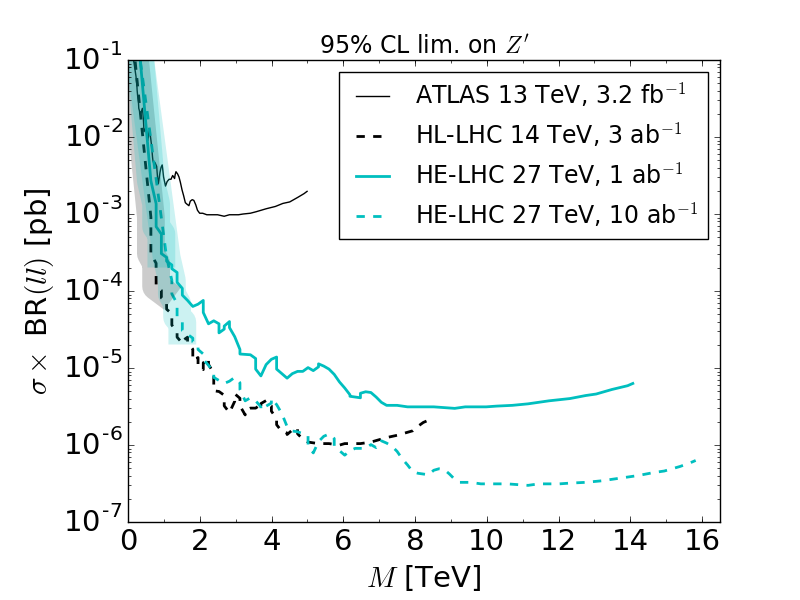}
\includegraphics[width=0.39 \textwidth]{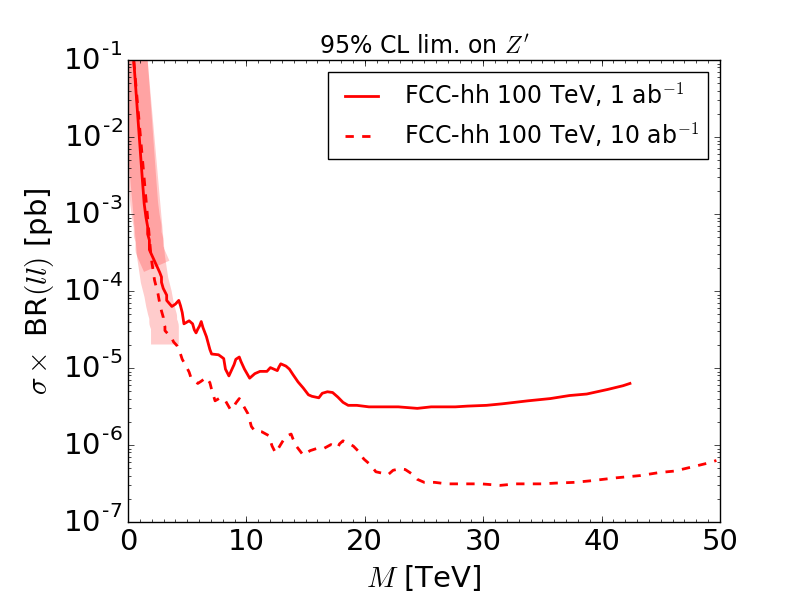}
\caption{Projected 95\% CL limits on di-muon final states at future colliders described in the legends. The conservative extrapolation method underestimates the actual limit at low masses, as indicated by the shaded region.}
\label{fig:minimalZprimeexclusion}
\end{figure}

We begin with the $Z^\prime$ case. Fig.~\ref{fig:minimalZprimeexclusion} shows the 95\% CL limit on the cross-section times branching ratio (in pb) for the di-muon final state. On the left plot the ATLAS limit for 13 TeV with 3.2 fb$^{-1}$ is denoted by a solid black line, and the limits extrapolated to HL-LHC and HE-LHC are represented by dashed black and cyan lines, respectively. On the right are the corresponding extrapolated limits for FCC-hh in red. The solid (dashed) cyan and red lines are for 1 (10) ab$^{-1}$ of integrated luminosity. 

\begin{figure}
\centering
\includegraphics[width=0.39 \textwidth]{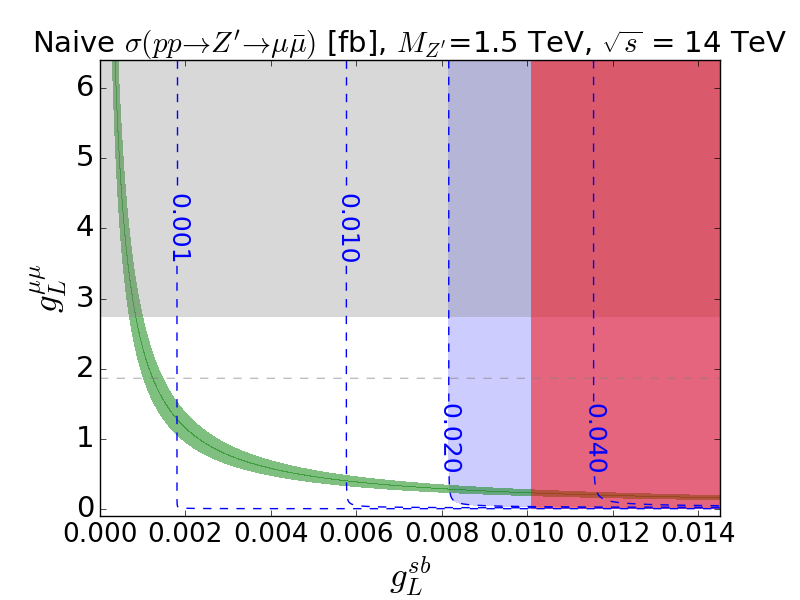}
\includegraphics[width=0.39 \textwidth]{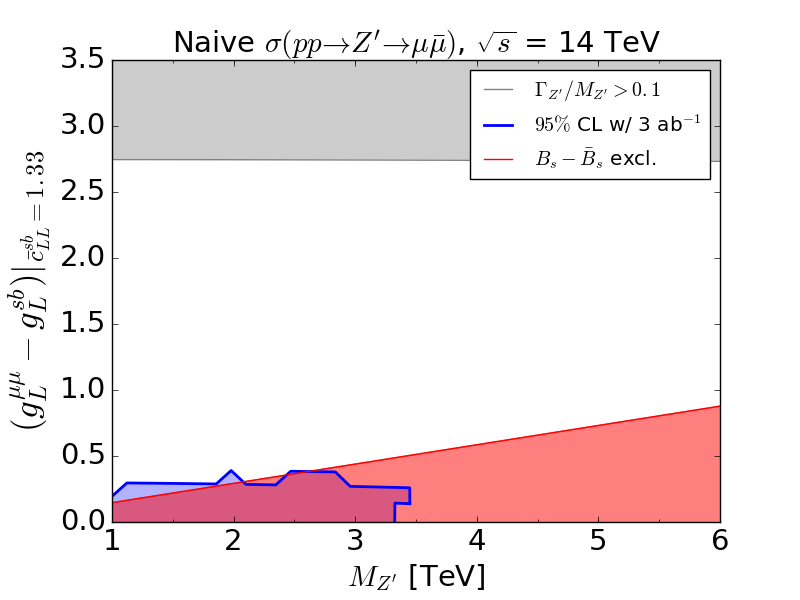}
\caption{Parameter space that explains the anomaly (shaded green on the left, whole region on the right) with HL-LHC coverage shown in blue and excluded by $B_s - \bar{B}_s$ mixing in red. The narrow width assumption breaks down in the grey region. }
\label{fig:14TeV}
\end{figure}

We may now apply these projected limits to our ``na\"{i}ve'' $Z^\prime$ model defined by coupling only to $sb$ and $\mu\mu$. As an example, the left plot of Fig.~\ref{fig:14TeV}  shows dashed blue contours of the cross-section (in units of fb) for $pp \to Z^\prime \to \mu^+\mu^-$ with  $M_{Z^\prime} = 1.5$ TeV at the HL-LHC in the plane of the $Z^\prime$ coupling strength, $g_{sb}$ vs $g_{\mu\mu}$. The parameter space compatible with explaining the $B$ anomalies lies along the green band. The 95\% CL reach is shaded in blue. We also see that increasing $g_{sb}$ beyond a certain point gives too large a contribution to $B_s - \bar{B}_s$ mixing, shaded in red. The grey region is where the $Z^\prime$ width is larger than 10\% of its mass and the narrow width approximation breaks down.

\begin{figure}
\centering
\includegraphics[width=0.39 \textwidth]{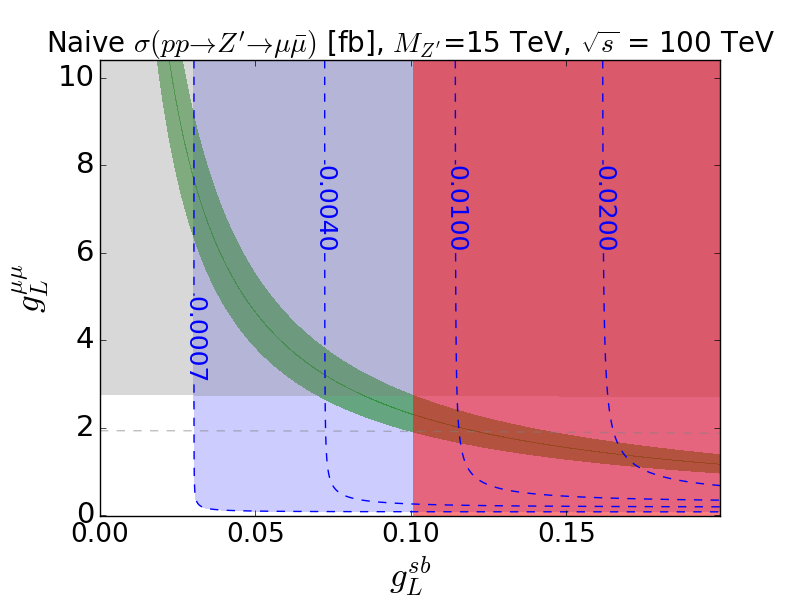}
\includegraphics[width=0.39 \textwidth]{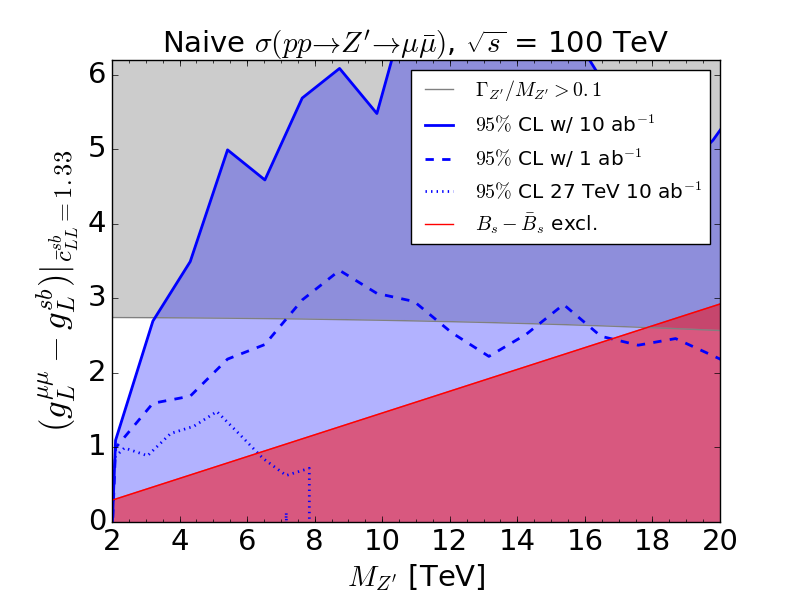}
\caption{As in Fig.~\ref{fig:14TeV} but for 100 TeV FCC-hh (and 27 TeV HE-LHC in dotted blue on the right).}
\label{fig:100TeV}
\end{figure}

The right plot in Fig.~\ref{fig:14TeV} summarises the coverage at the HL-LHC as a function of mass on the horizontal axis, with the vertical axis going along the anomaly-compatible green region for each mass. The colour-coding of the shaded regions are the same as described above, and we see that there is virtually no sensitivity to the large region of available parameter space. The HL-LHC is therefore not guaranteed to find a $Z^\prime$ if it is indeed the source of the anomalies. 

However, Fig.~\ref{fig:100TeV} shows the corresponding plots for a 100 TeV FCC-hh, with an example $M_{Z^\prime} = 15$ TeV plot on the left and the coverage summary plot on the right. The 27 TeV HE-LHC is also shown in dotted lines for comparison. While the na\"{i}ve model may still evade searches at HE-LHC, there is complete coverage for narrow width $Z^\prime$'s at FCC-hh (we note that the low mass regions are underestimated by the extrapolation method).

\begin{figure}
\centering
\includegraphics[width=0.39 \textwidth]{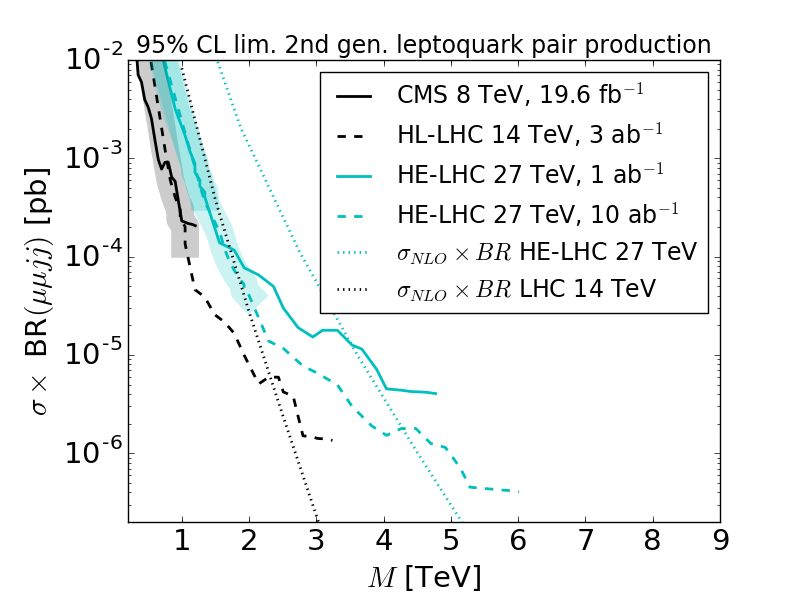}
\includegraphics[width=0.39 \textwidth]{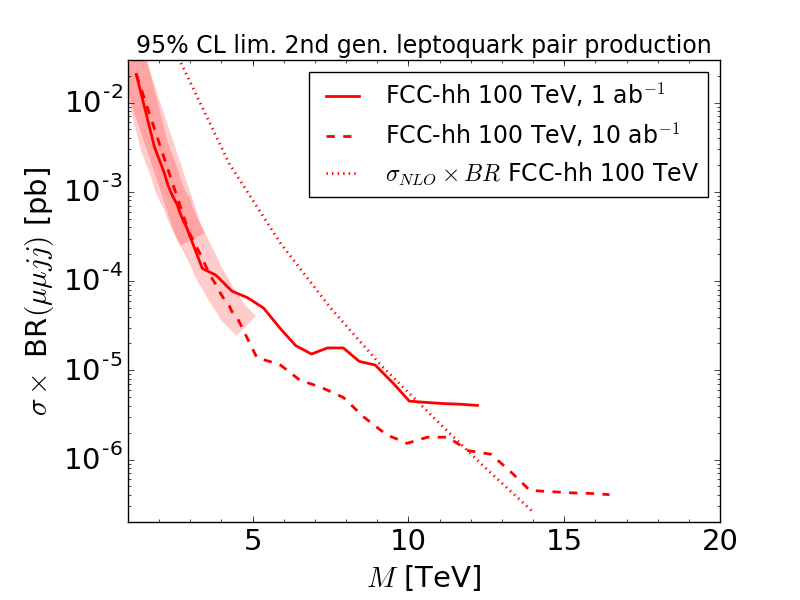}
\includegraphics[width=0.39 \textwidth]{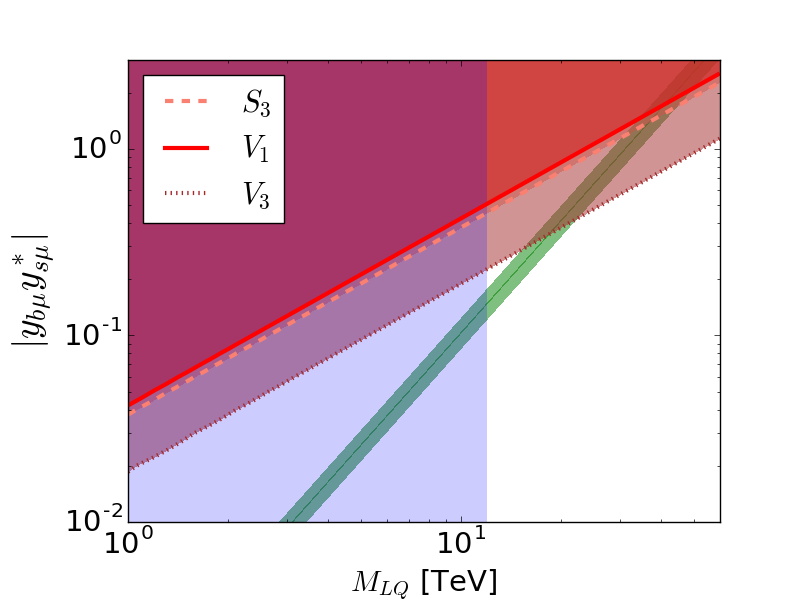}
\caption{95\% CL limits as described in Fig.~\ref{fig:minimalZprimeexclusion} but for leptoquark pair production. The available parameter space compatible with the anomaly in green is displayed on a log-log scale in the bottom plot, with the red region excluded by $B_s - \bar{B}_s$ mixing.} 
\label{fig:secondgenLQpairprodexclusion}
\end{figure}

We now turn to the case of leptoquarks. The 95\% CL limit for the $\mu\mu jj$ final state by CMS at 8 TeV is shown in solid black in the left plot of Fig.~\ref{fig:secondgenLQpairprodexclusion}. The extrapolation to HL-LHC, HE-LHC, and FCC-hh are as described above with the same colour coding. In this case the pair production of scalar leptoquarks is model-independent as it depends only on their coupling to gluons~\footnote{The vector leptoquark case is more model-dependent but limits are typically stronger for $\mathcal{O}(1)$ couplings. There are also model-dependent single leptoquark production limits~\cite{AGY}. }---the cross-section is therefore plotted directly on top of the limit curves, with masses excluded below their intersection. We see that the sensitivity to ``na\"{i}ve'' scalar leptoquarks reaches masses around 2.5, 4.5, and 12 TeV for HL-LHC, HE-LHC, and FCC-hh, respectively. However, the lower plot of Fig.~\ref{fig:secondgenLQpairprodexclusion} shows that anomaly-compatible leptoquarks may have masses up to 40 TeV before being excluded by $B_s - \bar{B}_s$ mixing.

\section{Conclusion}

In lieu of a general no-lose theorem, we may make the following conditional statement: 

{\it
If the neutral current $B$ anomalies are due to a narrow width $Z^\prime$ there is a good but not complete sensitivity at the HE-LHC, whereas it is guaranteed to be discovered by the FCC-hh; and if leptoquarks are responsible, then a discovery is only guaranteed for masses below 4.5 and 12 TeV at HE-LHC and FCC-hh respectively. 
}

The caveat is that ``guarantee'' is only meant in the sense that it is extremely unlikely (though not impossible) that nature will conspire to hide further the already contrived models considered here. On the contrary, these na\"{i}ve models are overly pessimistic and difficult to realise consistently~\footnote{It is, in fact, impossible to have a na\"{i}ve $Z^\prime$ model without additional couplings.}: for example, going from gauge to mass eigenstates generally induces further couplings, as in the so-called ``$33\mu\mu$'' $Z^\prime$ model which already has complete coverage at HE-LHC~\cite{AGY}. The situation for leptoquarks will also be more optimistic. 

Further refinements to these estimates can be made, and the sensitivity to more realistic benchmark scenarios can help determine the optimal strategy for targeting new physics. Even if the anomalies vanish, it is interesting to investigate the interplay between direct and indirect searches. On the other hand, should the anomalies be confirmed, it will be the start of a long, but tremendously exciting programme.

\section*{Acknowledgments}

I thank Ben Allanach and Ben Gripaios for their collaboration, and the organisers of the Rencontres de Moriond for inviting me to contribute this talk. I am supported by a Research Fellowship from Gonville and Caius College and partially supported by STFC consolidated grant ST/P000681/1.

\end{document}